\documentclass[12pt]{article}
\usepackage{amssymb,amsmath}

\begin{document}

\begin{center}
\null\vspace{2cm}
{\large {\bf Hawking Radiation via Tunneling from hot NUT-Kerr-Newman-Kasuya Spacetime}}\\
\vspace{2cm} M. Hossain Ali \footnote{\it Department of Applied
Mathematics, Rajshahi University, Rajshahi-6205, Bangladesh.}
\footnote{\it The Abdus Salam International Centre for Theoretical
Physics, Strada Costiera 11,\\ 34014 Trieste, Italy.
E-mail: $m_-hossain_-ali_-bd@yahoo.com$}\\

\end{center}
\vspace{3cm}
\centerline{\bf Abstract}
\baselineskip=18pt
\bigskip

We study the Hawking thermal spectrum in dragging
coordinate system and the tunneling radiation characteristics of 
hot NUT-Kerr-Newman-Kasuya spacetime. The tunneling rates at the event
and cosmological horizon are found to be related to the change of
Bekenstein-Hawking entropy. The radiation spectrum is not
pure thermal and thus there is a correction to the
Hawking thermal spectrum.
\vspace{0.5cm}

\vfill

\newpage

\section{Introduction}\label{sec1}

The signifying discovery of Stephen Hawking
\cite{one,two,three} that black hole radiates thermally raised
a disturbing and difficult problem, regarding information during
black hole evaporation, called as the information loss paradox.
Hawking's result also implies the loss of unitarity, or even the
breakdown of quantum mechanics \cite{four}, that is, the pure
quantum state is disintegrated to the mixture. In the language of
Quantum Field Theory, the ingoing state is the pure state, but the
outgoing is the mixture. Besides, Hawking regarded that the black
hole radiation is created by pair of particles via tunneling from
the black hole horizon as a result of the vacuum fluctuation. So
there does exist a tunneling process, but the created mechanism of
the tunneling barrier is unclear in this theory. The related
references do not use the language of quantum tunneling method to
discuss Hawking radiation, and hence it is not the quantum tunneling
method. In order to derive the radiant spectrum from the black hole horizon, one must solve the two difficulties: first, the formed mechanism of the potential hill; and secondly, the elimination of the coordinate singularity.

Recently, Hawking changed his opinion regarding the information loss
paradox and argued that information can indeed get out of the black
hole \cite{five}. Hawking's this argument is partly based on the
Parikh-Wilczek's recent work \cite{six,seven,eight} that treats
the Hawking radiation as tunneling process and applies the
semi-classical method to present the actual radiation as not exactly
thermal but subtle correction to the Hawking thermal spectrum.

Parikh-Wilczek, combined with the above reasons, proposed a
semi-classical quantum tunneling model that implemented Hawking
radiation as a tunneling process. This model was actually initiated
by Kraus and Wilczek (\cite{nine}, \cite{ten}) and Keski-Vakkuri and
Kraus \cite{eleven}, and was developed by Parikh and Wilczek with
considerable success by carrying out a dynamical treatment of black
hole geometry. More specially, they took into account the effects of
a positive energy matter shell propagating outwards through the
horizon of the Schwarzschild and Reissner-Nordstr\"om black holes,
and incorporated the self-gravitation correction of the radiation
(\cite{seven}, \cite{eight}, \cite{twelve}). In this theory, the
self-gravitation action among the particles creates the tunneling
barrier with turning points at the location of the black hole
horizon before and after the particle with energy emission. The
derived result implies that the radiant spectrum is not thermal, but is consistent with the underlying unitary theorem. This framework is so successful that it also satisfies the first law of black hole thermodynamic. Following this method, the radiation from spherically symmetric AdS black hole \cite{thirteen} and de Sitter cosmological horizon \cite{fourteen} were studied. Zhang and Zhao extended this method from spherical state to the
general axisymmetric Kerr and Kerr-Newman black holes, and even investigated more general massive and charged tunneling \cite{fifteen,sixteen,seventeen,eighteen}. More recently, Yang et al.
\cite{nineteen} studied the Hawking radiation as tunneling from
stationary axisymmetric Kerr-Newman-de Sitter black holes. They all
got a satisfying result. In addition, there were a lot of works
\cite{twenty,twenty one,twenty two,twenty three,twenty four,twenty five,twenty six,twenty seven,twenty eight,twenty nine,thirty,thirty one,thirty two,thirty three} in the last several years that
have attracted the interest of the scientific community. All of
these works indicate that the true Hawking radiate spectra are not
pure thermal when the self-gravitation is taken into account.

In this paper we apply Kraus-Parikh-Wilzcek's method to investigate
the Hawking radiation as tunneling from stationary axisymmetric
Kerr-Newman black hole spacetime in the de Sitter universe endowed
with NUT (magnetic mass) and magnetic monopole parameters, the
metric of which can be written as
\begin{equation}
\textrm{d}s^2=\frac{\Sigma}{\Delta_\theta}\textrm{d}\theta^2+\frac{\Sigma}{
\Delta_r}\textrm{d}r^2+\frac{\Delta_\theta\,\textrm{sin}^2\theta}{
\Sigma}\left(a\,\textrm{d}t_{HK}-\frac{\rho}{\Xi}\textrm{d}\varphi
\right)^2-\frac{\Delta_r}{\Sigma}\left(\textrm{d}t_{HK}-\frac{
\mathcal A}{\Xi}\textrm{d}\varphi\right)^2\label{eq1},
\end{equation}
where
\begin{eqnarray}
\Sigma&=&r^2+(n+a\,
\textrm{cos}\theta)^2,\hspace{0.5cm}\Delta_\theta=1+\frac{a^2}{\ell^2
}\textrm{cos}^2\theta,\hspace{0.5cm}\ell^2=\frac{3}{\Lambda},\nonumber\\
\Delta_r&=&\rho\left[1-\frac{1}{\ell^2}(r^2+5n^2)\right]-2(Mr+n^2)
+q_e^2+q_m^2,\nonumber\\
\rho&=&r^2+a^2+n^2,\hspace{0.5cm}\Xi=1+\frac{a^2}{\ell^2},\hspace{0.5cm}\mathcal A
=a\,\textrm{sin}^2\theta-2n\,\textrm{cos}\theta,\label{eq2}
\end{eqnarray}
$t_{HK}$ being the coordinate time of the spacetime. Beside the
cosmological parameter $\Lambda$, the metric (\ref{eq1}) possesses
five parameters: $M$ the mass parameter, $a$ the angular momentum
per unit mass parameter, $n$ the NUT (magnetic mass) parameter,
$q_e$ the electric charge parameter, and $q_m$ the magnetic monopole
parameter.

The metric (\ref{eq1}) solves the Einstein-Maxwell field equations
with an electromagnetic vector potential
\begin{equation}
A=-\frac{q_e\,r}{\sqrt{\Sigma\Delta_r}}e^0-\frac{q_m\textrm{
cos}\theta}{\sqrt{\Sigma\Delta_\theta}\,\textrm{sin}\theta}e^3\label{eq3},
\end{equation}
and an associated field strength tensor given by
\begin{eqnarray}
F&=&-\frac{1}{\Sigma^2}\left[q_e\,(r^2-a^2\textrm{cos}^2\theta)
+2q_m\,ra\,\textrm{cos}\theta\right]e^0\wedge e^1\nonumber\\
&&+\frac{1}{\Sigma^2}\left[q_m\,(r^2-a^2\textrm{cos}^2\theta)
-2q_e\,ra\,\textrm{cos}\theta\right]e^2\wedge e^3,\label{eq4}
\end{eqnarray}
where we have defined the vierbein field
\begin{eqnarray}
e^0&=&\sqrt{\frac{\Delta_r}{\Sigma}}\left(\textrm{d}t_{HK}-\frac{
\mathcal A}{\Xi}\textrm{d}\varphi\right),\hspace{0.5cm}e^1=\sqrt{\frac{
\Sigma}{\Delta_r}}\,\textrm{d}r,\nonumber\\
e^2&=&\sqrt{\frac{\Sigma}{\Delta_\theta}}\,\textrm{d}\theta,\hspace{0.5cm}e^3
=\sqrt{\frac{\Delta_\theta}{\Sigma}}\,\textrm{sin}\theta\left(a\,
\textrm{d}t_{HK}-\frac{\rho}{\Xi
}\textrm{d}\varphi\right)\label{eq5}.
\end{eqnarray}

We call the spacetime described by the metric (\ref{eq1}) a 
hot NUT-Kerr-Newman-Kasuya (or, for brevity, H-NUT-KN-K) spacetime, since the de Sitter spacetime has been interpreted as being hot \cite{thirty
four}. There is a renewed interest in the cosmological parameter as
it is found to be present in the inflationary scenario of the early
universe. In this scenario the universe undergoes a stage where it
is geometrically similar to de Sitter space \cite{thirty five}.
Among other things inflation has led to the cold dark matter. If the
cold dark matter theory proves correct, it would shed light on the
unification of forces (\cite{thirty six}, \cite{thirty seven}). The
monopole hypothesis was propounded by Dirac relatively long ago. The
ingenious suggestion by Dirac that magnetic monopole does exist was
neglected due to the failure to detect such object. However, in
recent years, the development of gauge theories has shed new light
on it. Moreover, the H-NUT-KN-K spacetime includes, among others,
the physically interesting black hole spacetimes as well as the NUT
spacetime which is sometimes considered as unphysical. The curious
properties of the NUT spacetime induced Misner \cite{thirty eight}
to consider it \lq \lq as a counter example to almost anything\rq
\rq . This spacetime plays a significant role in exhibiting the type
of effects that can arise in strong gravitational fields.

If we set $\ell\rightarrow\infty,\,a=q_e=q_m=0$ in Eq.(\ref{eq1}),
it then results the NUT metric which is singular along the axis of
symmetry $\theta=0$ and $\theta=\pi$. Because of the axial
singularities the metric admits different physical interpretations.
Misner \cite{thirty nine} introduced a periodic time coordinate to
remove the singularity, but this makes the metric an uninteresting
particle-like solution. To avoid a periodic time coordinate, Bonnor
\cite{fourty} removed the singularity at $\theta=0$ and related the
singularity at $\theta=\pi$ to a semiinfinite massless source of
angular momentum along the axis of symmetry. This is analogous to
representing the magnetic monopole in electromagnetic theory by
semiinfinite solenoid \cite{fourty one}. The singularity along
$z$-axis is analogous to the Dirac string.

McGuire and Ruffini \cite{fourty two} suggested that the spaces
endowed with the NUT parameter should never be directly physically
interpreted. To make a physically reasonable solution Ahmed
\cite{fourty three} used Bonnor's interpretation of the NUT
parameter, i.e., the NUT parameter $n$ is due to the strength of the
physical singularity on $\theta=\pi$, and further considered that
$n=a$. That means, the angular momentum of the mass $M$ and the
angular momentum of massless rod coalesce, and in this case, the
metric (\ref{eq1}) gives a new black hole solution which poses to
solve an outstanding problem of thermodynamics and black hole
physics. In view of all the above considerations the work of this paper is
interesting.

We organize the paper as follows. In section \ref{sec2}
we derive the horizons and the infinite red-shift surface for the
H-NUT-KN-K spacetime. We obtain Hawking thermal spectrum
in dragging coordinate system from Klein-Gordon equation in section
\ref{sec3}. Although the infinite red-shift surface and the horizons
are coincident with each other in the dragging coordinate system,
there still exists a coordinate singularity at the horizon of the
spacetime and it brings us inconvenience to investigate the
tunneling behavior across the horizon of the spacetime. In section
\ref{sec4} we remove the coordinate singularity by expressing the
metric in the Painlev\'e coordinate system and then investigate the
tunneling radiation characteristics of the H-NUT-KN-K spacetime.
Finally, in section \ref{sec5} we present our concluding remarks.

\section{Horizons and Infinite Red-shift Surface of H-NUT-KN-H Spacetime}\label{sec2}
The null surface equation $g^{\mu\nu}\partial_\mu f\partial_\nu f=0$
gives
\begin{equation}
r^4+(a^2+6n^2-\ell^2)r^2+2M\ell^2r-\left\{(a^2-n^2+q_e^2
+q_m^2)\ell^2-5n^2(a^2+n^2)\right\}=0\label{eq6}.
\end{equation}
Its roots locate the horizons of the H-NUT-KN-K spacetime. Let us
write it in the form
\begin{equation}
x^4+\alpha x^2+\beta x+\gamma=0\label{eq7}.
\end{equation}
Under the conditions
\begin{eqnarray}
&\alpha<0,\hspace{0.5cm}\beta>0,\hspace{0.5cm}\gamma<0,\hspace{0.5cm
}\alpha^2+12\gamma>0,&\nonumber\\
&(\alpha^2+12\gamma)^3>(\alpha^3-36\alpha\gamma+\frac{27}{2}\beta^2)^2,&\label{eq8}
\end{eqnarray}
equation (\ref{eq7}) has four real roots: three of which are
positive $x^+_1,\,x^+_2,\,x^+_3$, and one which is negative $x^-$,
\begin{eqnarray}
&&x^+_1+x^+_2+x^+_3+x^-=0;\label{eq9}\\
&&x^+_1=-w_1+w_2+w_3,\nonumber\\
&&x^+_2=w_1-w_2+w_3,\nonumber\\
&&x^+_3=w_1+w_2-w_3,\label{eq10}
\end{eqnarray}
where
\begin{eqnarray}
&&w_1=\left[-\frac{1}{6}\alpha+\frac{1}{6}\sqrt{(\alpha^2+12\gamma)
}\,\textrm{cos}\left(\frac{1}{3}\psi\right)\right]^{1/2},\nonumber\\
&&w_2=\left[-\frac{1}{6}\alpha-\frac{1}{6}\sqrt{(\alpha^2+12\gamma)
}\,\textrm{cos}\left(\frac{1}{3}\psi+\frac{1}{3}\pi\right)\right]^{1/2},\nonumber\\
&&w_3=\left[-\frac{1}{6}\alpha-\frac{1}{6}\sqrt{(\alpha^2+12\gamma)
}\,\textrm{cos}\left(\frac{1}{3}\psi-\frac{1}{3}\pi\right)\right]^{1/2},\label{eq11}\\
&&\textrm{cos}\psi=\frac{\alpha^3-36\alpha\gamma+\frac{27}{2}\beta^2}{
(\alpha^2+12\gamma)^{3/2}}\label{eq12}.
\end{eqnarray}

All the four roots of (\ref{eq6}) can be evaluated according to the
formulae (\ref{eq9})-(\ref{eq12}). We denote the negative root by
$r_-$, and the remaining three positive real roots which represent
the horizons of the H-NUT-KN-K spacetime, namely, the inner, outer
(event), and cosmological horizon, respectively, are given by
\begin{eqnarray}
&&r_0=-t_1+t_2+t_3,\nonumber\\
&&r_H=t_1-t_2+t_3,\nonumber\\
&&r_C=t_1+t_2-t_3\label{eq13},
\end{eqnarray}
where
\begin{eqnarray}
&&t_1=\left[\frac{1}{6}(\ell^2-6n^2-a^2)+\frac{1}{6}\sqrt{\{(\ell^2-6n^2-a^2)^2-\digamma\}}
\,\textrm{cos}\left(\frac{1}{3}\psi\right)\right]^{1/2},\nonumber\\
&&t_2=\left\{\frac{1}{6}(\ell^2-6n^2-a^2)-\frac{1}{6}\sqrt{\{(\ell^2-6n^2-a^2)^2-\digamma\}
}\,\textrm{cos}\left(\frac{1}{3}\psi+\frac{1}{3}\pi\right)\right\}^{1/2},\nonumber\\
&&t_3=\left\{\frac{1}{6}(\ell^2-6n^2-a^2)-\frac{1}{6}\sqrt{\{(\ell^2-6n^2-a^2)^2-\digamma\}
}\,\textrm{cos}\left(\frac{1}{3}\psi-\frac{1}{3}\pi\right)\right\}^{1/2},\label{eq14}\\
&&\textrm{cos}\psi=-\frac{(\ell^2-6n^2-a^2)\{(\ell^2-6n^2-a^2)^2+3\digamma\}-54M^2\ell^4
}{\{(\ell^2-6n^2-a^2)^2-\digamma\}^{3/2}},\nonumber\\
&&\digamma=12\left\{(q_e^2+q_m^2-n^2+a^2)\ell^2-5n^2(a^2+n^2)\right\},\label{eq15}
\end{eqnarray}
under the conditions
\begin{eqnarray}
\{(\ell^2-6n^2-a^2)^2-\digamma\}^3&>&\{(\ell^2-6n^2-a^2)^3
+3(\ell^2-6n^2-a^2)\digamma-54M^2\ell^4\}^2\nonumber\\
(\ell^2-6n^2-a^2)&>&0\label{eq16}.
\end{eqnarray}

For the constant time-slice and $r=r_H$, the metric (\ref{eq1})
reduces to
\begin{equation}
\textrm{d}\sigma^2=\frac{\Sigma}{\Delta_\theta}\textrm{d}\theta^2+\frac{\Delta_\theta\,
\textrm{sin}^2\theta}{\Xi^2\Sigma}(r_H^2+a^2+n^2)^2\textrm{d}\varphi^2\label{eq17},
\end{equation}
the determinant of this two-dimensional line element is
\begin{equation}
g=\frac{\textrm{sin}^2\theta}{\Xi^2}(r_H^2+a^2+n^2)^2\label{eq18}.
\end{equation}
Then the area of the event horizon can be expressed as
\begin{equation}
A_H=\int \textrm{d} A^\prime=\int \sqrt{g}\,\textrm{d}\theta
\textrm{d}\varphi=\frac{4\pi}{\Xi}(r_H^2+a^2+n^2)\label{eq19},
\end{equation}
and that of the cosmological horizon as
\begin{equation}
A_C=\frac{4\pi}{\Xi}(r_C^2+a^2+n^2)\label{eq20}.
\end{equation}

For the infinite red-shift surface: $g_{00}=0$, we obtain
\begin{equation}
\Delta_r-\Delta_\theta\,a^2\textrm{sin}^2\theta=0\label{eq21}.
\end{equation}
It is obvious that the infinite red-shift surface and the event
horizon of the spacetime are not coincident with each other, an energy layer exists between them. So the geometrical opticts limit cannot be used here. We therefore carry on dragging coordinate transformation and let
\begin{equation}
\dot{\varphi}=\frac{\textrm{d}\varphi}{\textrm{d}t_{HK}}
=-\frac{g_{03}}{g_{33}}=\Omega\label{eq22}.
\end{equation}
The spacetime line element (\ref{eq1}) in the dragging coordinate system is
\begin{equation}
\textrm{d}s^2=\hat{g}_{00}\textrm{d}t_{HK}^2
+\frac{\Sigma}{\Delta_r}\textrm{d}r^2
+\frac{\Sigma}{\Delta_\theta}\textrm{d}\theta^2\label{eq23},
\end{equation}
where
\begin{equation}
\hat{g}_{00}=g_{00}-\frac{g_{03}^2}{g_{33}}
=-\frac{\Delta_\theta\Delta_r(\rho-a\mathcal A)^2\textrm{sin}^2\theta}
{\Sigma(\Delta_\theta\rho^2\textrm{sin}^2\theta-\Delta_r\mathcal A^2)}\label{eq24}.
\end{equation}

In fact, the line element (\ref{eq23}) represents a three-dimensional hypersurface in the four-dimensional H-NUT-KN-K spacetime. Evidently, the infinite red-shift surface is coincident with the
horizons of the spacetime in the dragging coordinate system when
$\hat{g}_{00}$ vanishes.

\section{Hawking Thermal Spectrum of H-NUT-KN-K Spacetime
in Dragging Coordinate System}\label{sec3}

In this section, for the sake of simplicity, we investigate the
Hawking thermal radiation spectrum of uncharged particles. The
Klein-Gordon equation for uncharged particles in the curved
spacetime can be expressed in the form
\begin{equation}
\frac{1}{\sqrt{-g}}\frac{\partial}{\partial
x^\mu}\left(\sqrt{-g}\,g^{\mu\nu}\frac{\partial}{\partial
x^\nu}\Phi\right)=\mu^2\Phi\label{eq25}.
\end{equation}
With
\begin{equation}
\Phi=e^{-\textrm{i}\omega
t_{HK}}R(r)\Theta(\theta)e^{\textrm{i}m\varphi}\label{eq26},
\end{equation}
$g^{\mu\nu}$ from Eq.(\ref{eq23}), and considering the dragging
coordinate transformation (\ref{eq22}), we obtain the following
expression
\begin{eqnarray}
\frac{\textrm{d}^2R(r)}{\textrm{d}r^2}&+&\frac{1}{g^{11}}
\left(\frac{g^{11}}{\sqrt{-g}}\frac{\partial}{\partial
r}\sqrt{-g}+\frac{\partial g^{11}}{\partial r}\right)
\frac{\textrm{d}R(r)}{\textrm{d}r}+\frac{1}{g^{11}}\frac{R(r)}
{\Theta (\theta)}\{G(r,\theta)\}\nonumber\\
&=&\frac{1}{g^{11}}\left[\mu^2+\left(\omega+m\frac{g_{03}}{g_{33}
}\right)^2\hat{g}^{00}\right]R(r)\label{eq27},
\end{eqnarray}
where
\begin{equation}
G(r,\theta)=\frac{\Delta_\theta}{\Sigma}
\frac{\textrm{d}^2\Theta(\theta)}{\textrm{d}\theta^2}
+\frac{1}{\sqrt{-g}}\frac{\partial}{\partial\theta
}\left(\sqrt{-g}\,g^{22}\right)
\frac{\textrm{d}\Theta(\theta)}{\textrm{d}\theta}\label{eq28}.
\end{equation}

Introducing the tortoise coordinate
\begin{equation}
r_\ast=\frac{1}{2\kappa_H}\textrm{ln}(r-r_H)\label{eq29},
\end{equation}
we have
\begin{eqnarray}
\frac{\textrm{d}^2R(r)}{\textrm{d}r_\ast^2}&-&
2\kappa_H\frac{\textrm{d}R(r)}{\textrm{d}r_\ast}
+2\kappa_H(r-r_H)\left(\frac{1}{\sqrt{-g}}\frac{\partial\sqrt{-g}}
{\partial r}+\frac{1}{g^{11}}\frac{\partial g^{11}} {\partial
r}\right)\frac{\textrm{d}R(r)}{\textrm{d}r_\ast}\nonumber\\
&+&\frac{4\kappa_H^2(r-r_H)^2}{g^{11}}\frac{R(r)}
{\Theta (\theta)}\{G(r,\theta)\}\nonumber\\
&=&\frac{4\kappa_H^2(r-r_H)^2}{g^{11}}
\left[\mu^2+\left(\omega+m\frac{g_{03}}{g_{33}
}\right)^2\hat{g}^{00}\right]R(r)\label{eq30},
\end{eqnarray}
where
\begin{equation}
\kappa_H=\frac{1}{2\ell^2(r_H^2+a^2+n^2)}(r_H-r_-)(r_H-r_0)(r_C-r_H)\label{eq31}
\end{equation}
is the surface gravity of the event horizon. In the vicinity of the
event horizon, i.e., when $r\rightarrow r_H$, one could find that
\begin{equation}
\frac{4\kappa_H^2(r-r_H)^2}{g^{11}}
\left[\mu^2+\left(\omega+m\frac{g_{03}}{g_{33}
}\right)^2\hat{g}^{00}\right]R(r)\equiv-(\omega-m\Omega_H)^2R(r)\label{eq32}.
\end{equation}
Equation (\ref{eq30}) then can be put, near the horizon, in the
standard wave equation form:
\begin{equation}
\frac{\textrm{d}^2R(r)}{\textrm{d}r_\ast^2}+(\omega-\omega_0)^2R(r)=0\label{eq33},
\end{equation}
where $\omega_0=m\Omega_H=\frac{\Xi am}{r_H^2+a^2+n^2}$,
$\Omega_H$ being the dragging angular velocity at the event horizon.
Solving Eq.(\ref{eq33}) we obtain the ingoing and outgoing radial
wave functions for uncharged particles in the H-NUT-KN-K spacetime
as follows:
\begin{equation}
\Phi_{\textrm{in}}=e^{-\textrm{i}\omega\upsilon},\hspace{1.0cm}\Phi_{\textrm{out}}
=e^{-\textrm{i}\omega\upsilon}e^{2\textrm{i}(\omega-\omega_0)r_\ast}\label{eq34},
\end{equation}
where $\upsilon=t_{HK}+\frac{\omega-\omega_0}{\omega}r_\ast$ is the
advanced Eddington-Finkelstein coordinate. Near the event horizon,
$\Phi_{\textrm{out}}$ can be written as
\begin{equation}
\Phi_{\textrm{out}}=e^{-\textrm{i}\omega\upsilon}(r-r_H)^{
\textrm{i}(\omega-\omega_0)/\kappa_H}\label{eq35}.
\end{equation}
The $\Phi_{\textrm{out}}$ has a logarithm singularity. By analytical
continuation rotating $-\pi$ through the lower-half complex
$r$-plane, we have
\begin{equation}
(r-r_H)\rightarrow\mid r-r_H\mid
e^{-\textrm{i}\pi}=(r-r_H)e^{-\textrm{i}\pi}\label{eq36}.
\end{equation}

Damour and Ruffini's work \cite{fourty four} of generalizing the classical approach of barrier penetration to curved spaces endowed with future horizons, allows one to recover most directly the spectrum of the Hawking radiation. The existence of a spacelike Killing vector $\xi _t$ inside the horizon permits a classical particle as \lq \lq seen\rq \rq from infinity to reach a negative-energy state. In the quantum description, this phenomenon allows an antiparticle to reach positive-energy states, and these states can be tunneled out by a wave function \lq \lq over\rq \rq the horizon. This gives rise to the creation of a pair: one particle (positive energy) going out and one antiparticle (negative energy) falling back toward the singularity.

Using the Damour-Ruffini stretch method of analysis, and extending it to the inside of the event horizon, we obtain the spectrum of the Hawking radiation \cite{fourty four}
\begin{equation}
N_\omega=\frac{1}{e^{(\omega-\omega_0)/T}-1}=\frac{1}{e^{\alpha_HA_H}-1}\label{eq37},
\end{equation}
where
\begin{equation}
T=\frac{\kappa_H}{2\pi},\hspace{1.0cm}
\alpha_H=\frac{3(\omega-\omega_0)\Xi\ell^2} {[-\digamma
r_H^{-1}+18M\ell^2+6(a^2+6n^2-\ell^2)r_H]}\label{eq38},
\end{equation}
$A_H$ being the area of the event horizon and $\digamma$ is given by
(\ref{eq15}). It is obvious from Eq.(\ref{eq37}) that the Hawking
radiation spectrum at the event horizon is related to the fixed area
of the event horizon.

In similar fashion, the Hawking radiation spectrum at the
cosmological horizon is given by
\begin{equation}
N_\omega=\frac{1}{e^{(\omega-\omega_0)/T}-1}=\frac{1}{e^{\alpha_CA_C}-1}\label{eq39},
\end{equation}
where
\begin{equation}
T=\frac{\kappa_C}{2\pi},\hspace{1.0cm}
\alpha_C=\frac{3(\omega-\omega_0)\Xi\ell^2} {[-\digamma
r_C^{-1}+18M\ell^2+6(a^2+6n^2-\ell^2)r_C]}\label{eq40},
\end{equation}
Equation (\ref{eq39}) shows that the derived Hawking radiation
spectrum at the cosmological horizon is also related to the fixed
area of the cosmological horizon, $A_C$.

Thus the thermal property of the H-NUT-KN-K spacetime can be derived
in the dragging coordinate system. The expressions (\ref{eq37}) and
(\ref{eq39}) are based on the fixed background spacetime. In fact,
the horizons change with the emission, and the background
spacetime is dynamical.

\section{Painlev\'e Coordinate Transformation and Tunneling Process
of Uncharged Particles from H-NUT-KN-K Spacetime}\label{sec4}
\subsection{Painlev\'e-H-NUT-KN-K Metric}\label{subsec4.1}
In the dragging coordinate system, the infinite red-shift surface
coincides with the horizons, but still there is a coordinate
singularity at the horizon of the spacetime, which brings us
inconvenience to investigate the tunneling process across the
horizon. So we perform general Painlev\'e coordinate transformation
\cite{fourty five} to eliminate the coordinate singularity from the
metric (\ref{eq23}) as follows. We write
\begin{equation}
\textrm{d}t_{HK}=\textrm{d}t+F(r, \theta)\textrm{d}r+G(r,
\theta)\textrm{d}\theta\label{eq41},
\end{equation}
where $F(r, \theta)$ and $G(r, \theta)$ are two functions to be
determined about $r$, $\theta$, and satisfy the integrability
condition $\partial_\theta F(r, \theta)=\partial_r G(r, \theta)$.
Substituting Eq.(\ref{eq41}) into the metric (\ref{eq23}), and
ordering the derived constant-time slice of the spacetime flat
Euclidean in the radial, we obtain the metric in the general
Painlev\'e coordinate system:
\begin{eqnarray}
\textrm{d}s^2&=&\hat{g}_{00}\textrm{d}t^2
+\textrm{d}r^2\pm2\sqrt{\hat{g}_{00}(1-g_{11})}\,
\textrm{d}t\textrm{d}r+\left[\hat{g}_{00}G^2(r,\theta)
+g_{22}\right]\textrm{d}\theta^2\nonumber\\
&&+2\sqrt{\hat{g}_{00}(1-g_{11})}\,G(r,
\theta)\textrm{d}r\textrm{d}\theta+2\hat{g}_{00}G(r,
\theta)\textrm{d}t\textrm{d}\theta\label{eq42},
\end{eqnarray}
where the positive sign (+) represents the spacetime of the outgoing
particle, and the negative sign (-) denotes the metric of the
ingoing particle. Flat Euclidean space consideration gives
\begin{equation}
F(r, \theta)=\pm\sqrt{(1-g_{11})/\hat{g}_{00}}\label{eq43}.
\end{equation}
According to Landau's condition of coordinate clock synchronization
\cite{fourty six}
\begin{equation}
\frac{\partial}{\partial
x^i}\left(-\frac{g_{0j}}{\hat{g}_{00}}\right)=\frac{\partial}{\partial
x^j}\left(-\frac{g_{0i}}{\hat{g}_{00}}\right)\label{eq44},
\end{equation}
from which, for the metric (\ref{eq42}), we also have
\begin{equation}
\frac{\partial F(r, \theta)}{\partial\theta} =\frac{\partial G(r,
\theta)}{\partial r}\label{eq45}.
\end{equation}
Thus the Painlev\'e-H-NUT-KN-K metric (\ref{eq42}) satisfies the
Landau's condition of the coordinate clock synchronization. In
addition, there are many other superior features: firstly, the
metric is regular at the horizons; secondly, the infinite red-shift
surface and the horizons are coincident with each other; thirdly,
spacetime is stationary; and fourthly, constant-time slices are just
flat Euclidean space in radial. All of these properties are
advantageous for us to study the Hawking thermal spectrum via
tunneling.

In order to investigate the tunneling behavior of the uncharged
particles from the horizon, we first evaluate the radial, null
geodesics. Since the tunneling processes take place near the event
horizon, we may consider a particle tunneling across the event
horizon as an ellipsoid shell and think that the particle should
still be an ellipsoid shell during the tunneling process, i.e., the
particle does not have motion in the $\theta$-direction. Under these
assumptions ($\textrm{d}s=0=\textrm{d}\theta$), the radial, null
geodesics followed by uncharged particles are, from (\ref{eq42}),
\begin{equation}
\dot{r}=\frac{dr}{dt}=\frac{\sqrt{\Delta_\theta}(\rho-a\mathcal A)
\textrm{sin}\theta(\pm\sqrt{\Sigma}-\sqrt{\Sigma-\Delta_r})}
{\sqrt{\Sigma(\Delta_\theta\rho^2\textrm{sin}^2\theta-
\Delta_r\mathcal A^2)}}\label{eq46},
\end{equation}
where the plus (minus) sign indicates the outgoing (ingoing)
geodesics, under the implicit assumption that time $t$ increases
towards the future.

\subsection{Tunneling Process}\label{subsec4.2}
We now turn to discuss the Hawking radiation of uncharged particles
as a semi-classical tunneling process across the barrier which is
created just by the outgoing particle itself. We adopt the picture
of a pair of virtual particles spontaneously created just inside the
horizon. The positive energy virtual particle can tunnel out and
materialize as a real particle escaping classically to infinity, and
the negative energy anti-particle is absorbed by the H-NUT-KN-K
spacetime, resulting in a decrease in the mass and angular momentum
of the H-NUT-KN-K spacetime. We consider the particle as an
ellipsoid shell of energy $\omega$ and angular momentum $a\omega$.
If the particle's self-gravitation is taken into account,
Eqs.(\ref{eq13}), (\ref{eq42}), and (\ref{eq46}) should be modified.
To guarantee the conservation of energy and angular momentum, we fix
the total mass and total angular momentum of the whole of the
spacetime (i.e., the H-NUT-KN-K spacetime plus the outside
spacetime) but allow the the H-NUT-KN-K spacetime's mass and angular
momentum to fluctuate. When the particle is tunneled out as an
ellipsoid shell of energy $\omega$ and angular momentum $a\omega$,
then the mass and angular momentum of the H-NUT-KN-K spacetime will
be replaced by $(M-\omega)$ and $a(M-\omega)$, respectively.
Meanwhile, the event horizon will shrink, we refer to the cases pre-
and post-shrinking as two turning points of potential barrier. The
distance between the two turning points is the width of potential
barrier and decided by the energy of outgoing particle. In this
critical situation, Eqs.(\ref{eq13}), (\ref{eq42}), and (\ref{eq46})
will be modified by replacing the mass parameter $M$ with
$(M-\omega)$, and the shell of energy will move along the modified
null geodesic in the radial direction
\begin{equation}
\dot{r}=\frac{\sqrt{\Delta_\theta}(\rho-a\mathcal A)
\textrm{sin}\theta(\pm\sqrt{\Sigma}-\sqrt{\Sigma-\Delta_r^\prime})}
{\sqrt{\Sigma(\Delta_\theta\rho^2\textrm{sin}^2\theta-
\Delta_r^\prime\mathcal A^2)}}\label{eq47},
\end{equation}
where
\[
\Delta_r^\prime=(r^2+a^2+n^2)\left[1-\frac{1}{\ell^2}(r^2+5n^2)\right]
-2\left\{(M-\omega)r+n^2\right\}+q^2_e+q^2_m
\]
is the horizon equation after the emission of the particle with
energy $\omega$.

In the WKB approximation, the tunneling probability for an outgoing
positive energy particle can be expressed in terms of the imaginary
part of the action as
\begin{equation}
\Gamma\sim e^{-2\textrm{Im}\,S}\label{eq48}.
\end{equation}
At this point, it should be noticed that the coordinate $\varphi$
does not appear in the metric in the dragging coordinate system.
That is, $\varphi$ is an ignorable coordinate in the Lagrangian
function, $L(r, \dot{r}, \varphi, \dot{\varphi},t)$. To eliminate
this degree of freedom completely, the imaginary part of the action
should be written as
\begin{eqnarray}
\textrm{Im}\,S&=&\textrm{Im}\int_{t_i}^{t_f}(p_r\dot{r}-
p_\varphi\dot{\varphi})\textrm{d}t\nonumber\\
&=&\textrm{Im}\int_{r_i}^{r_f}\left[\int_{(0,\,0)}^{(p_r,\,p_\varphi)}
(\dot{r}\textrm{d}p_r^\prime-\dot{\varphi}
\textrm{d}p_\varphi^\prime)\right]\frac{\textrm{d}r}{\dot{r}}\label{eq49},
\end{eqnarray}
where $p_r$ and $p_\varphi$ are two canonical momenta conjugate to
$r$ and $\varphi$, respectively. The $r_i$ and $r_f$ are just inside
and outside the barrier at the event horizon through which the
particle tunnels.

We now remove the momentum in favor of energy by applying the
Hamilton's equations
\begin{eqnarray}
\dot{r}&=&\left.\frac{\textrm{d}H}{\textrm{d}p_r}\right|_{(r;\,
\varphi,\,p_\varphi)},\nonumber\\
\dot{\varphi}&=&\left.\frac{\textrm{d}H}{\textrm{d}p_\varphi}
\right|_{(\varphi;\,r,\,p_r)},
\hspace{0.5cm}\textrm{d}H|_{(\varphi;\,r,\,p_r)}=\Omega^\prime
\textrm{d}J\label{eq50},
\end{eqnarray}
where $H=\frac{M}{\Xi^2}$ is the total energy of the H-NUT-KN-K
spacetime, and when the particle of energy $\omega$ is propagating
from inside to outside the event horizon, then
$H=\frac{M-\omega}{\Xi^2}$,
$\textrm{d}H=-\frac{1}{\Xi^2}\textrm{d}\omega$, $p_\varphi=J$.

Substituting Eqs.(\ref{eq47}) and (\ref{eq50}) into Eq.(\ref{eq49}),
we obtain
\begin{eqnarray}
\textrm{Im}\,S&=&\textrm{Im}\int_{\frac{M}{\Xi^2}}^{\frac{M-\omega}{\Xi^2}
}\int_{r_i}^{r_f}\left(\frac{\textrm{d}H^\prime}{\dot{r}}
-\frac{\Omega^\prime\textrm{d}J^\prime}{\dot{r}}\right)\textrm{d}r
=\textrm{Im}\int_{0}^{\omega}\int_{r_i}^{r_f}-\frac{1}{\Xi^2}
\left(\frac{\textrm{d}\omega^\prime}{\dot{r}}
-\frac{a\Omega^\prime\textrm{d}\omega^\prime}{\dot{r}}\right)\textrm{d}r\nonumber\\
&=&\textrm{Im}\int_{0}^{\omega}\int_{r_i}^{r_f}-\frac{1}{\Xi^2}
\frac{\sqrt{(\Delta_\theta\rho^2\textrm{sin}^2\theta-\tilde{\Delta}_r^\prime\mathcal A^2)}\,
\left[\Sigma\sqrt{\Delta_\theta}+\sqrt{\Sigma\Delta_\theta(\Sigma-\tilde{\Delta}_r^\prime})\right]
}{\Delta_\theta(\rho-a\mathcal A)\textrm{sin}\theta\tilde{\Delta}_r^\prime}\nonumber\\
&&\times(1-a\Omega^\prime)\textrm{d}\omega^\prime\textrm{d}r\label{eq51},
\end{eqnarray}
where
\[
\tilde{\Delta}_r^\prime=(r^2+a^2+n^2)\left[1-\frac{1}{\ell^2}(r^2+5n^2)\right]
-2\{(M-\omega^\prime)r+n^2\}+q_e^2+q_m^2.
\]
There is a single pole in Eq.(\ref{eq51}) at the event horizon of
the H-NUT-KN-K spacetime after the particle emission. We can
evaluate the integral by deforming the contour around the pole, so
as to ensure that positive energy solution decay in time. In this
way, we finish firstly the $\omega^\prime$ integral and obtain the
result
\begin{equation}
\textrm{Im}\,S=\textrm{Im}\int_{r_i}^{r_f}-\frac{\pi
r\textrm{i}}{\Xi}\textrm{d}r=-\frac{\pi}{2\Xi}(r_f^2-r_i^2)\label{eq52}.
\end{equation}
In terms of the entropy expression
$S_{BH}=\pi(r_H^2+a^2+n^2)/\Xi$, the tunneling rate at the event
horizon can then be expressed, using Eq.(\ref{eq48}), as follows:
\begin{equation}
\Gamma\sim e^{-2\textrm{Im}\,S}=e^{\Delta S_{BH}}\label{eq53},
\end{equation}
where $\Delta S_{BH}=S_{BH}^\prime-S_{BH}$ is the difference of
Bekenstein-Hawking entropies of the H-NUT-KN-K spacetime before and
after the emission of the particle. From comparison of
Eqs.(\ref{eq53}) and (\ref{eq37}), we can learn that the tunneling
rate at the event horizon provides a correct modification to Hawking
radiation spectrum.

Let us now discuss the the Hawking radiation of the particle via
tunneling at the cosmological horizon. The particle is found
tunneled into the cosmological horizon differently from the
particle's tunneling behavior of the event horizon. When the
particle with energy $\omega$ tunnels into the cosmological horizon,
Eqs.(\ref{eq13}), (\ref{eq42}) and (\ref{eq46}) should have to
modify by replacing the mass parameter $M$ with $(M+\omega)$ after
taking the self-gravitation action into account. Thus, after
tunneling the particle with energy $\omega$ into the cosmological
horizon, the null radial geodesic takes the form
\begin{equation}
\dot{r}=\frac{\sqrt{\Delta_\theta}(\rho-a\mathcal A)
\textrm{sin}\theta(\pm\sqrt{\Sigma}-\sqrt{\Sigma-\Delta_r^{\prime\prime}})}
{\sqrt{\Sigma(\Delta_\theta\rho^2\textrm{sin}^2\theta-
\Delta_r^{\prime\prime}\mathcal A^2)}}\label{eq54},
\end{equation}
where
\[
\Delta_r^{\prime\prime}=(r^2+a^2+n^2)\left[1-\frac{1}{\ell^2}(r^2+5n^2)\right]
-2\left\{(M+\omega)r+n^2\right\}+q^2_e+q^2_m.
\]
Different from the event horizon, $H=-\frac{M}{\Xi^2}$ and
$H^\prime=-\frac{M+\omega}{\Xi^2}$ are the total energy of the
H-NUT-KN-K spacetime before and after the particle with energy
$\omega$ tunnels into. Then the imaginary part of the action at the
cosmological horizon can be written as
\begin{eqnarray}
\textrm{Im}\,S&=&\textrm{Im}\int_{-\frac{M}{\Xi^2}}^
{-\frac{M-\omega}{\Xi^2}
}\int_{r_{Ci}}^{r_{Cf}}\left(\frac{\textrm{d}H^\prime}{\dot{r}}
-\frac{\Omega^\prime\textrm{d}J^\prime}{\dot{r}}\right)\textrm{d}r
=\textrm{Im}\int_{0}^{\omega}\int_{r_{Ci}}^{r_{Cf}}-\frac{1}{\Xi^2}
\left(\frac{\textrm{d}\omega^\prime}{\dot{r}}
-\frac{a\Omega^\prime\textrm{d}\omega^\prime}{\dot{r}}\right)\textrm{d}r\nonumber\\
&=&\textrm{Im}\int_{0}^{\omega}\int_{r_{Ci}}^{r_{Cf}}-\frac{1}{\Xi^2}
\frac{\sqrt{(\Delta_\theta\rho^2\textrm{sin}^2\theta
-\tilde{\Delta}_r^{\prime\prime}\mathcal A^2)}\,
\left[\Sigma\sqrt{\Delta_\theta}+\sqrt{\Sigma\Delta_\theta(\Sigma
-\tilde{\Delta}_r^{\prime\prime})}\right]
}{\Delta_\theta(\rho-a\mathcal A)\textrm{sin}\theta\tilde{\Delta}_r^{\prime\prime}}\nonumber\\
&&\times(1-a\Omega^\prime)\textrm{d}\omega^\prime\textrm{d}r\label{eq55},
\end{eqnarray}
where
\[
\tilde{\Delta}_r^{\prime\prime}=(r^2+a^2+n^2)\left[1-\frac{1}{\ell^2}(r^2+5n^2)\right]
-2\{(M+\omega^\prime)r+n^2\}+q_e^2+q_m^2,
\]
$r_{Ci}$ and $r_{Cf}$ are the locations of the cosmological horizon
before and after the particle of energy $\omega$ tunneling into.
There exists a single pole at the cosmological horizon in
Eq.(\ref{eq55}). Finishing the $\omega^\prime$ integral firstly, we
obtain
\begin{equation}
\textrm{Im}\,S=\textrm{Im}\int_{r_{Ci}}^{r_{Cf}}-\frac{\pi
r\textrm{i}}{\Xi}\textrm{d}r=-\frac{\pi}{2\Xi}(r_{Cf}^2-r_{Ci}^2)\label{eq56}.
\end{equation}
Using the entropy expression $S_{CH}=\pi(r_C^2+a^2+n^2)/\Xi$,
the tunneling rate at the cosmological horizon is given by
\begin{equation}
\Gamma\sim e^{-2\textrm{Im}\,S}=e^{\Delta S_{CH}}\label{eq57},
\end{equation}
where $\Delta S_{CH}=S_{CH}^\prime-S_{CH}$ is the difference of
Bekenstein-Hawking entropies of the cosmological horizon before and
after the particle of energy $\omega$ is tunneling into. Comparing
Eqs.(\ref{eq57}) and (\ref{eq39}), we can find that the tunneling
rate at the cosmological horizon still provides a correct
modification to Hawking radiation spectrum.

\section{Concluding Remarks}\label{sec5}
The main concern of this paper has been exclusively the
investigation of the tunneling radiation characteristics of
uncharged particles from a more general spacetime, namely, the 
hot NUT-Kerr-Newman-Kasuya spacetime, by applying Kraus-Parikh-Wilczek's
semi-classical quantum tunneling method (\cite{seven}, \cite{eight},
\cite{twelve}). Our result is satisfactory.

In our study, we find that the tunneling rate at the
event/cosmological horizon is related to the change of
Bekenstein-Hawking entropy and the radiant spectrum is no
longer thermal after considering the H-NUT-KN-K spacetime
background as dynamical and incorporating the self-gravitation
effect of the emitted particles when the energy conservation and
angular momentum conservation are taken into account. Thus our study
is perfectly extending the Kraus-Parikh-Wilczek's semi-classical
tunneling framework in a more general spacetime, containing six
parameters: the mass $M$, angular momentum per unit mass $a$,
cosmological parameter $\Lambda$, NUT (magnetic mass) parameter $n$,
electric charge $q_e$ and magnetic monopole charge $q_m$.

In special cases, our result reduces to the Reissner-Norstr\"om
black hole case for $\ell\rightarrow\infty$, $a=0=n$, and to the
Schwarzschild black hole case for $\ell\rightarrow\infty$, $a=0=n$,
$q_e=0=q_m$, and supports the Parikh's result (\cite{seven},
\cite{eight}, \cite{twelve}).

For $n=0$ and $q_e^2+q_m^2=q^2$, our study gives the result of Yang
et al. \cite{nineteen} for the Kerr-Newman-de Sitter black hole.
Indeed, by suitably choosing the parameters of the spacetime, the
result of this paper can be specialized for all the interesting
black hole spacetimes, de Sitter spacetimes as well as the NUT
spacetime which has curious properties as discussed in the
introduction. Following Ahmed \cite{fourty three}, if one considers
the Bonnor's interpretation: the NUT parameter $n$ is due to the
strength of the physical singularity on $\theta=\pi$, and further
chooses $n=a$, then our study gives result for the interesting
coalescing black hole spacetime in the de Sitter universe. In
addition, our result can be directly extended to the anti-de Sitter
case by changing the sign of the cosmological parameter $\ell^2$ to
a negative one. In view of all of the above attractive features, the
study of this paper is interesting.

Using Kraus-Parikh-Wilzcek's method, Hawking radiation of charged
massive particles as a semi-classical tunneling process across the
horizons of the H-NUT-KN-K spacetime \cite{fourty seven} is interesting as well.

Hawking radiation from the event horizon of black hole is one of the
most important achievements  of quantum field theory in curved
spacetimes. In fact, due to Hawking evaporation classical general
relativity, statistical physics, and quantum field theory are
connected in quantum black hole physics. It is, therefore, generally
believed that the deep investigation of black hole physics would be
helpful to set up a satisfactory quantum theory of gravity.

\vspace{1.0cm}

\noindent
{\large\bf Acknowledgement}\\
I am thankful to SIDA as well as the Abdus Salam International
Centre for Theoretical Physics, Trieste, Italy, where this paper was
produced during my Associateship visit.

\end{document}